\documentclass[conference]{IEEEtran}
\IEEEoverridecommandlockouts
\newcommand{\etal}{\textit{et al.}}
\usepackage{cite}
\usepackage{balance}
\usepackage{amsmath,amssymb,amsfonts}
\usepackage{algorithmic}
\usepackage{graphicx}
\usepackage{textcomp}
\usepackage{xcolor}
\usepackage{subcaption}
\usepackage{multirow}
\usepackage{array}
\usepackage{ragged2e}
\usepackage{dblfloatfix}
\usepackage{placeins}
\def\BibTeX{{\rm B\kern-.05em{\sc i\kern-.025em b}\kern-.08em
    T\kern-.1667em\lower.7ex\hbox{E}\kern-.125emX}}
\begin{document}

\title{Motif-based morphology signatures for interpretable ECG screening and monitoring \\
{\footnotesize}
\thanks{NB and MV are funded by The Podium Institute of Sports Medicine and 
Technology.}
}

\author{\IEEEauthorblockN{Nivedita Bijlani}
\IEEEauthorblockA{\textit{
The Podium Institute for Sports Medicine and Technology} \\
\textit{University of Oxford}\\
Oxford, UK \\
0000-0001-9862-4732}
\and
\IEEEauthorblockN{Mauricio Villarroel}
\IEEEauthorblockA{\textit{
The Podium Institute for Sports Medicine and Technology} \\
\textit{University of Oxford}\\
Oxford, UK \\
0000-0003-4787-6053}
}

\maketitle

\begin{abstract}
Electrocardiography (ECG) remains central to cardiovascular screening, yet interpretation remains largely manual and episodic. Clinical practice relies on brief resting ECGs and, when required, long-duration ambulatory recordings, both generating data that require resource-intensive review. Consequently, subtle morphological changes or progressive drift preceding clinically apparent abnormalities may go unnoticed. We propose a motif-based framework that defines beat-aligned ECG motifs as interpretable cardiac signatures and quantifies morphological drift and deviation across short and long-term monitoring. Motifs are representative cardiac cycles capturing dominant morphology. We introduce three interpretable drift metrics: deviation from a normal sinus rhythm (NSR), deviation from a personalised baseline, and a motif instability index. Motifs are extracted by selecting beats that minimise Dynamic Time Warping (DTW) distance within fixed windows. We evaluate these metrics on short (PTB-XL) and long-duration (MIT-BIH Arrhythmia) ECG datasets. Interpretability is achieved through representative motif overlays and fiducial-based visualisations, enabling direct inspection of morphological changes. In MIT-BIH, the proposed metrics significantly separated predominantly normal from arrhythmic subjects ($p<0.01$). In PTB-XL, NSR deviation distinguished normal from abnormal ECGs across major diagnostic subtypes ($p<10^{-4}$, Cliff’s $\delta$ up to 0.93). ECG motifs provide an interpretable representation of cardiac morphology, supporting scalable longitudinal monitoring and early detection of morphology-driven change.
\end{abstract}


\section{Introduction}
Cardiovascular disease remains the leading cause of death globally, accounting for an estimated 17.9 million deaths each year~\cite{b1}. While age-standardised rates have stabilised, the absolute burden is projected to rise substantially, with cardiovascular prevalence increasing by 90\% and annual deaths reaching 35.6 million by 2050~\cite{b2}. Many cardiac pathologies, including arrhythmias, myocardial infarction, and sudden cardiac arrest, begin with subtle, progressive changes in cardiac electrophysiology. Electrocardiography (ECG) is routinely used as the first-line tool for cardiac screening in at-risk individuals and following clinical events. Standard practice relies on brief resting ECGs, supplemented by longer ambulatory monitoring when symptoms are infrequent or transient. In athletic populations, where sudden cardiac death has become a growing clinical concern~\cite{b3}, regular ECG screening is often mandated as a prerequisite for participation. Continuous ECG monitoring through wearable devices across rest, stress, sleep, and daily activity—is also increasingly used to capture patterns associated with early or evolving cardiac pathology. 

Despite its widespread use, ECG interpretation remains largely manual and episodic, requiring systematic review by trained clinicians and substantial expert time~\cite{b4}. Both short clinical ECGs and long-duration ambulatory recordings generate substantial data volumes that are resource-intensive to review. As a result, subtle morphological changes, gradual drift, or emerging deviations preceding clinical manifestation may be overlooked. This challenge is particularly acute in young athletes, where ECG interpretation is complicated by training-related adaptation and directly informs decisions around eligibility to play, training load, and recovery~\cite{b5}. There is a clear need for methods that improve the efficiency of ECG screening and enable early intervention. This requires identifying dominant and recurring ECG morphological patterns as interpretable ``cardiac signatures'' to support prioritisation and informed clinical review.

Motifs can provide a simple and efficient framework for constructing cardiac signatures from ECG data. Motifs are short time series that represent recurring or approximately repeated patterns~\cite{b6}\cite{b7}\cite{b8}. By encoding ECG recordings into representative single-cardiac-cycle motifs, dominant morphology can be identified. Motifs are extracted within short temporal windows and can be applied consistently to both brief clinical ECGs and long-duration ambulatory recordings. When applied across longer recordings, motifs can capture changes in  morphology, enabling rapid identification of patterns and deviations while supporting expert review.

We hypothesise that changes in motif composition, drift, and stability provide informative markers of emerging cardiac abnormality. Tracking motif behaviour over time enables the construction of personalised cardiac signatures that reflect both stable physiology and evolving deviations. Such signatures offer a principled way to support ECG pre-screening and prioritisation, while remaining directly interpretable at the waveform level.

Motif-based pattern discovery has been widely studied in time-series analysis, where recurring subsequences provide compact representations of temporal structure~\cite{b9}\cite{b10}\cite{b11}. However, motif-based approaches remain under-explored in biophysical signals, where extracted patterns must correspond to clinically meaningful physiological structure and change. Schäfer and Leser~\cite{b12} introduced k-Motiflets, defining motifs as compact sets of mutually similar subsequences, and demonstrated detection of recurring physiological patterns in ECG and electroencephalogram (EEG) signals. However, motifs are treated as generic subsequences, rather than events explicitly anchored to the cardiac cycle. Their temporal evolution is not modelled, limiting interpretability for longitudinal ECG analysis. Sivaraks and Ratanamahatana~\cite{b13} proposed a motif-based framework for ECG artifact and anomaly detection using time-domain similarity measures, including Dynamic Time Warping (DTW). While high detection accuracy was achieved, motif discovery is driven by frequency changes and relies on fixed reference templates, limiting its ability to capture evolving morphology and intra-subject variability. Ramezani~\cite{b14} introduced a self-supervised ECG motif framework based on spectral segmentation and symbolic representations, achieving strong arrhythmia classification performance. However, motifs are not cardiac-cycle aligned, and morphology is encoded via frequency-domain abstractions rather than beat-level structure. Also, motif evolution is not explicitly modelled longitudinally.

Motif-based approaches have also been applied to a range of biophysical time series, including EEG and other physiological signals. Kraljevska~\etal~\cite{b15} applied motif-based pattern discovery to EEG recordings to predict treatment response in major depressive disorder, while Germain~\etal~\cite{b16} proposed a persistence-based framework for identifying recurring patterns using topological features. Uudeberg~\etal~\cite{b17} introduced the in-phase matrix profile (pMP) to quantify signal self-similarity in EEG, and Dvirnas~\etal~\cite{b18} demonstrated the use of Matrix Profile for scalable subsequence matching in biophysical data.

Existing motif-based approaches largely rely on generic subsequence matching, heuristic parameter choices, or templates optimised for detecting specific abnormalities rather than characterising ECG morphology. As a result, motifs are often not physiologically interpretable at the cardiac-cycle level, limiting their clinical utility for understanding waveform structure, tracking morphological change, or supporting informed ECG interpretation. To address this gap, we define beat-aligned ECG motifs as interpretable cardiac signatures relative to normal sinus rhythm and use their temporal behaviour to quantify morphological drift and stability. 

Our contributions are:
\begin{enumerate}
\item We formalise beat-aligned ECG motifs as representative cardiac cycles that capture dominant cardiac rhythm morphology.
\item We introduce motif-based drift metrics to quantify deviation, stability, and variability of ECG morphology over time.
\item We demonstrate that motif-based metrics separate normal and abnormal morphology across both short clinical and long-duration ambulatory ECG datasets.
\end{enumerate}

\section{Methods}
\begin{figure*}[t]
\centering
\includegraphics[width=1.0\textwidth]{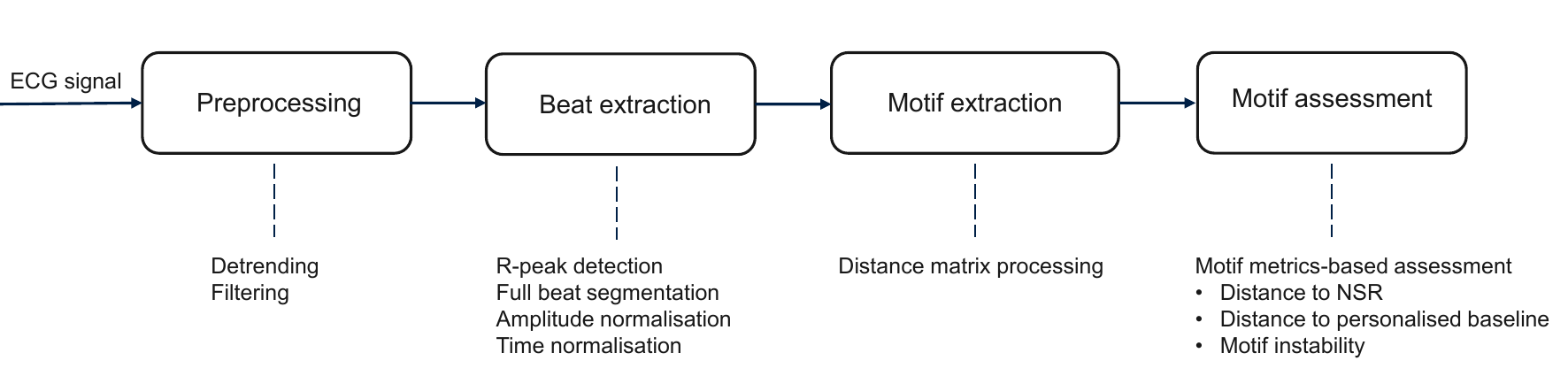}
\caption{Overview of the proposed motif-based ECG analysis pipeline, including preprocessing, beat extraction and normalisation, motif extraction, and motif-based assessment.}
\label{fig:workflow}
\end{figure*}
An overview of the proposed workflow from preprocessing to motif-based assessment is shown in Fig.~\ref{fig:workflow}.

\subsection{Datasets}
This study utilises two large, publicly available, expert-annotated ECG datasets: the MIT-BIH Arrhythmia Database~\cite{b19} and PTB-XL~\cite{b20}. MIT-BIH consists of 48 half-hour ambulatory ECG recordings from 47 subjects with beat-level annotations, while PTB-XL comprises 21,837 10\,s clinical ECGs spanning multiple diagnostic classes. From PTB-XL, we focussed on abnormal recordings corresponding to conditions primarily characterised by morphological ECG alterations, namely conduction disturbances, ST-segment and T-wave (ST/T) changes, hypertrophy, and myocardial infarction. The normal cohort comprised only recordings labelled exclusively as \textit{Normal}. For MIT-BIH, recordings were categorised as predominantly normal if $\geq$90\% of the 10\,s analysis windows contained only normal sinus beats. All remaining recordings were treated as abnormal. The datasets are summarised in Table~\ref{tab:dataset_summary}. All analyses were performed at the record level. Lead I was used for PTB-XL, while the modified limb lead II (MLII) was used for MIT-BIH recordings.

\subsection{ECG preprocessing}
Each ECG recording is pre-processed to remove baseline drift, high-frequency noise, and power-line interference. This is achieved using a 4th-order low-pass Butterworth filter (40\,Hz) and a 2nd-order high-pass Butterworth filter (0.05\,Hz), with additional notch filtering at 60\,Hz for MIT-BIH and 50\,Hz for PTB-XL to reflect dataset-specific acquisition environments. R-peaks are detected using \texttt{nk.ecg\_findpeaks}
(method = ``neurokit'')~\cite{b21}.
Cardiac cycles are R-peak aligned and temporally normalised to a fixed-length representation of 650 samples spanning the full R--R interval, from the preceding to the subsequent R-peak, ensuring inclusion of both depolarisation and repolarisation phases. This preserves relative waveform morphology while removing heart-rate dependence, yielding a heart-rate--invariant representation focused on ECG morphology. The fixed length was chosen to provide sufficient temporal resolution for fiducial structure. Temporal normalisation is applied uniformly across beats for shape comparison under DTW; frequency content is not explicitly interpreted. Each cycle is then mean-centred and scaled to unit amplitude, enabling baseline- and scale-invariant comparison across beats and subjects. Beats are screened for quality based on baseline wander, high-frequency noise, and clipping. Motifs are extracted from valid beats within a single-lead ECG waveform to ensure consistency across datasets and avoid variability introduced by lead-specific morphology. 

\subsection{Motif extraction}
\label{subsec:motif_extraction}
We briefly define time-series motifs as per~\cite{b22}. Let a univariate time series be denoted by $T = \{t_1, t_2, \ldots, t_n\}$. 
A subsequence $T_{i,L}$ is a contiguous segment of $T$ of length $L$, starting at index $i$. A motif is defined as the pair of distinct subsequences of length $L$ whose distance is minimal among all possible subsequence pairs in $T$. 
Formally, $T_{a,L}$ and $T_{b,L}$ form a motif if
\[
\mathrm{dist}(T_{a,L}, T_{b,L}) \le \mathrm{dist}(T_{i,L}, T_{j,L}), 
\quad \forall\, i \neq j,
\]
where $\mathrm{dist}(\cdot,\cdot)$ denotes a chosen distance measure. In the ECG setting, motif similarity is evaluated using Dynamic Time Warping (DTW) as the distance measure. 
Given two cardiac cycles $X = \{x_1, \ldots, x_m\}$ and $Y = \{y_1, \ldots, y_n\}$, DTW computes the minimum cumulative alignment cost over all admissible warping paths:
\begin{equation}
\mathrm{DTW}(X,Y) = \min_{W} \sum_{(i,j)\in W} \lVert x_i - y_j \rVert,
\end{equation}
where $W$ denotes a warping path that preserves temporal ordering and continuity, and $\lVert \cdot \rVert$ is the pointwise distance. To enable fair comparison across beats and windows, we use a normalised DTW distance defined as the cumulative DTW cost divided by the length of the optimal warping path and scaled to a fixed reference beat length. This yields a length-independent distance measure that reflects average per-sample morphological deviation.

Motifs are extracted within fixed, non-overlapping 10\,s windows. For short 10\,s ECG recordings, this yields a single representative motif per recording. For long-duration recordings, motifs are extracted sequentially across windows, forming a motif trajectory that summarises the temporal evolution of recurring cardiac morphology. In addition to motifs, we retain the corresponding discords within each window, defined as the cardiac cycle most dissimilar from all others under DTW. In the ECG context, discords capture extreme morphological deviation from the dominant pattern and support downstream interpretation of emerging abnormality.

\subsection{Normal sinus rhythm (NSR) template definition}
\begin{figure}[tb]
\centering
\includegraphics[scale=0.8]{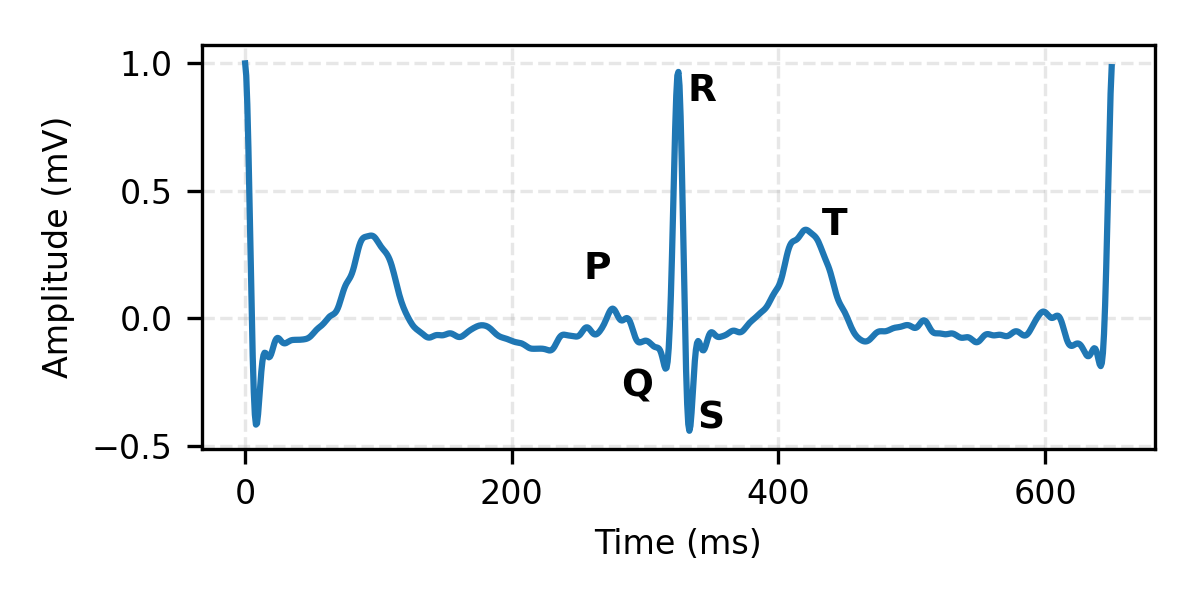}
\caption{Normal sinus rhythm (NSR) motif template derived from normal-labelled male ECGs (20--30 years, Lead I) in PTB-XL. Cardiac cycles are R-peak aligned and temporally normalised to 650 samples over the full R--R interval, preserving waveform morphology while removing heart-rate dependence. Cycles are mean-centred and scaled to unit amplitude for baseline- and scale-invariant comparison.}
\label{fig:NSR template}
\end{figure}

To construct a reference normal sinus rhythm (NSR) template, ECG recordings from healthy male subjects aged 20--30 years with no labelled cardiac pathology are selected from PTB-XL to minimise age- and disease-related morphological variability. Motifs are extracted from these recordings as described in Subsection~\ref{subsec:motif_extraction}, yielding a set of representative cardiac-cycle waveforms \(\{\mathbf{y}_k\}_{k=1}^{K}\). The NSR template motif, shown in Fig.~\ref{fig:NSR template}, is then defined as the DTW medoid of this motif set:
\begin{equation}
\mathbf{y}_{\mathrm{NSR}} = \arg\min_{k \in \{1,\dots,K\}} \sum_{j=1}^{K} \mathrm{DTW}(\mathbf{y}_k,\mathbf{y}_j),
\end{equation}
corresponding to the waveform that minimises the total DTW distance to all other motifs in the normal reference set.

\subsection{Motif-based morphological metrics}
We define three motif-based metrics to quantify deviation and temporal instability of ECG morphology: \newline

\paragraph{NSR Morphological Deviation Index (NMDI)}
Deviation from NSR morphology is quantified as the DTW distance between an ECG motif $\mathbf{y}_m$ and the population-level NSR template $\mathbf{y}_{\mathrm{NSR}}$:
\begin{equation}
D_{\mathrm{NSR}} = \mathrm{DTW}(\mathbf{y}_m, \mathbf{y}_{\mathrm{NSR}}).
\end{equation}
This metric measures deviation from the morphology encoded by the NSR reference template.\newline

\paragraph{Personalised Morphological Deviation Index (PMDI)}
Deviation from an individual’s typical morphology is quantified relative to a personalised baseline motif $\mathbf{y}_{\mathrm{base}}$, which for long-term ECGs is defined as the motif extracted from the first analysis window. This provides a simple initial estimate of typical morphology.\newline

\paragraph{Motif Instability Index (MII)}
Temporal instability of morphology is quantified as the mean DTW distance between motifs $\mathbf{y}_m$ extracted from consecutive windows:
\begin{equation}
I = \frac{1}{W-1} \sum_{w=1}^{W-1} \mathrm{DTW}(\mathbf{y}_m^{(w)}, \mathbf{y}_m^{(w+1)}),
\end{equation}
where $W$ denotes the number of windows. Higher values indicate increased morphological drift over time.

\subsection{Quantitative assessment of motif-based ECG separation}
We assessed the ability of motif-derived deviation metrics to capture morphological differences between normal and abnormal ECG recordings. Record-level distributions of motif deviation from the NSR template (NMDI) were compared. Additionally, for longer recordings ($>$10\,s), drift from personalised baselines (PMDI) and morphological instability (MII) were compared. Group-wise differences were assessed using the Mann--Whitney U test, with statistical significance reported via $p$-values. Effect sizes were quantified using Cliff’s delta ($\delta$), a non-parametric measure of the magnitude and direction of separation between abnormal and normal ECG morphology.

\subsection{Visual interpretation of morphological deviations}
Morphological differences between individual ECGs and the NSR template are decomposed into amplitude and interval deviations at key ECG fiducials and visualised using summary plots to support physiological interpretation.

\section{Experiments and Results}
All experiments were run on a 64-bit Windows 11 system with an Intel Ultra 7 165H CPU (3.8\,GHz) and 32\,GB RAM.

\begin{table}[tb]
\centering
\caption{ECG datasets used for motif-based analysis}
\label{tab:dataset_summary}
\small
\setlength{\tabcolsep}{4pt}
\renewcommand{\arraystretch}{1.07}
\begin{tabular}{|l|>{\RaggedRight\arraybackslash}p{3.3cm}|c|c|}
\hline
\textbf{Dataset} & \textbf{Class} & 
\textbf{\begin{tabular}{c}
Records/ \\
Subjects
\end{tabular}} & 
\textbf{\begin{tabular}{c}
10\,s Analysis \\
Windows
\end{tabular}} \\
\hline
\multirow{5}{*}{PTB-XL} 
& Normal & 7,172 & 7,172 \\
\cline{2-4}
& Abnormal (Conduction Disturbances) & 4,483 & 4,483 \\
\cline{2-4}
& Abnormal (ST/T Changes) & 5,202 & 5,202 \\
\cline{2-4}
& Abnormal (Hypertrophy) & 2,644 & 2,644 \\
\cline{2-4}
& Abnormal (Myocardial Infarction) & 5,468 & 5,468 \\
\hline
\multirow{2}{*}{MIT-BIH}
& Normal ($\geq$90\% normal beat windows) & 10 & 1,800 \\
\cline{2-4}
& Abnormal & 38 & 6,840 \\
\hline
\end{tabular}
\end{table}

\begin{figure*}[!t]
\centering
\includegraphics[width=\textwidth]{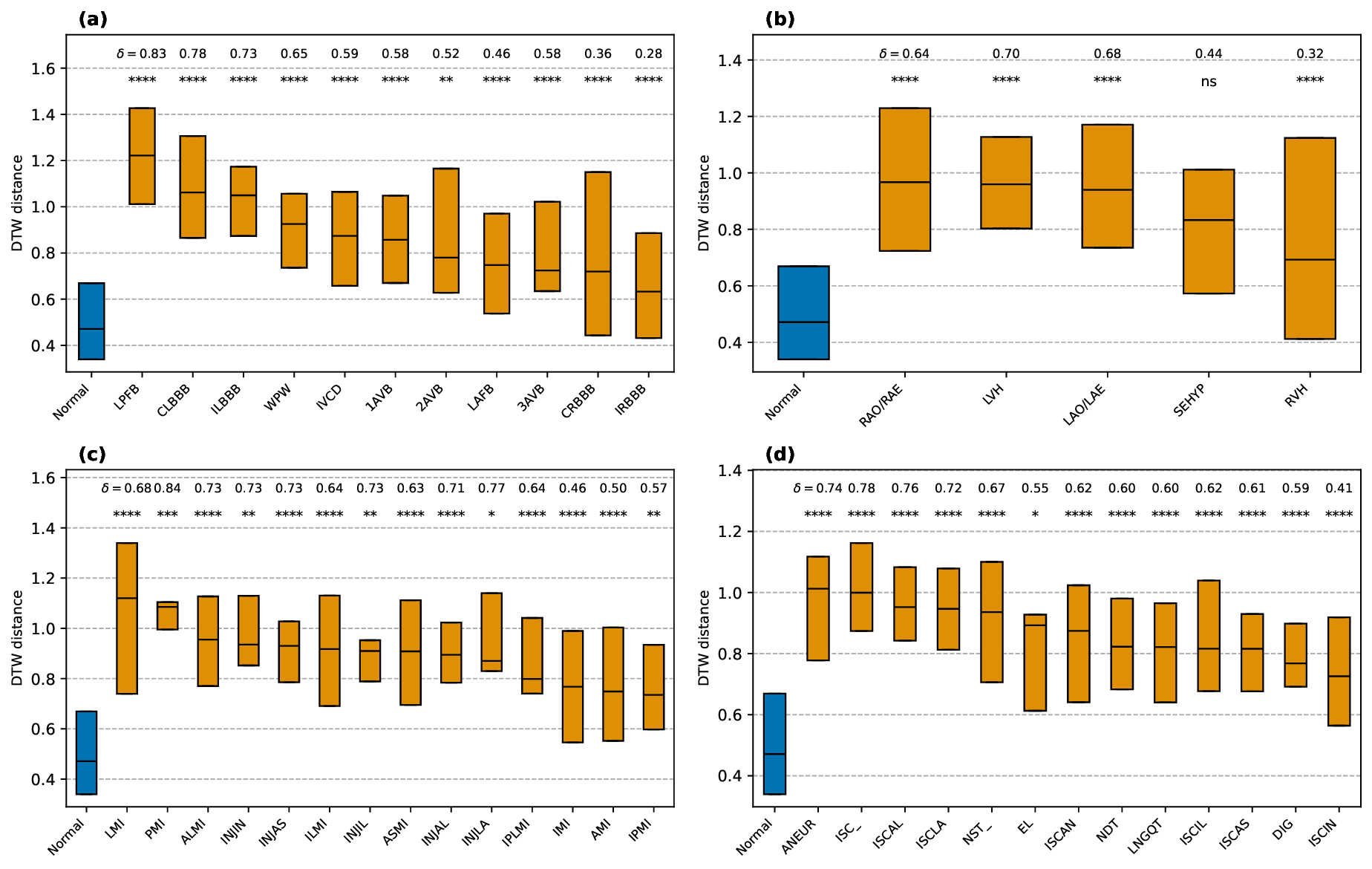}
\caption{Distribution of motif distances to the NSR template motif (NMDI) for normal ECGs and abnormal subtypes in PTB-XL. Boxplots compare normal recordings with (a) conduction disturbances (CD), (b) hypertrophy (HYP), (c) myocardial infarction (MI), and (d) ST/T changes. Effect sizes (Cliff’s $\delta$) and statistical significance are reported above each comparison, highlighting morphological deviation from normal sinus rhythm.}
\label{fig:PTB-XL results}
\end{figure*}

\subsection{Quantitative validation of motif-based metrics}
\begin{figure}[htb]
\centering
\includegraphics[scale=0.73]{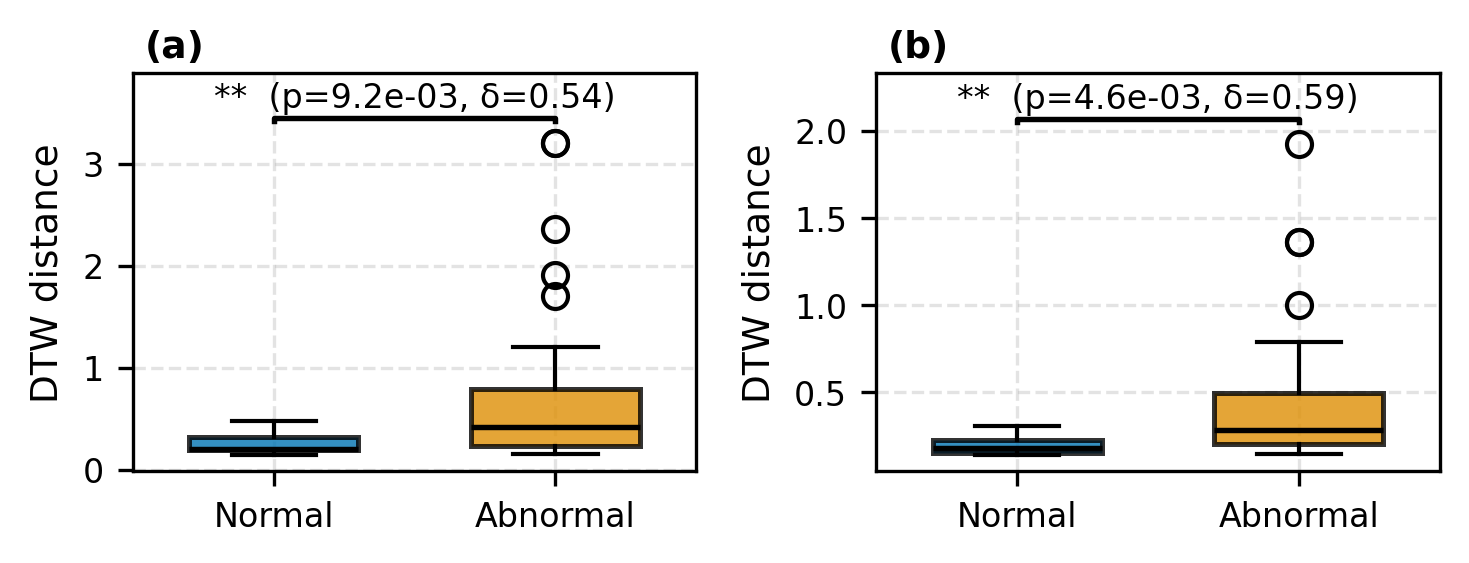}
\caption{(a) Distribution of motif distances to the personalised baseline motif  (PMDI) for normal and abnormal ECGs. (b) Distribution of mean motif distances between consecutive windows (MII) for normal and abnormal ECGs.}
\label{fig:MIT BIH-motif deviation boxplots}
\end{figure}

In the PTB-XL dataset, the NMDI was significantly higher in abnormal ECGs across conduction disturbance, hypertrophy, myocardial infarction, and ST/T change subgroups compared to normal recordings (typical $p < 0.001$, Cliff’s $\delta = 0.28$--$0.89$), as shown in Fig.~\ref{fig:PTB-XL results}. In the MIT-BIH database, strong separation was observed for the PMDI ($p = 9.2 \times 10^{-3}$, $\delta = 0.54$) and MII ($p = 4.6 \times 10^{-3}$, $\delta = 0.59$), as shown in Fig.~\ref{fig:MIT BIH-motif deviation boxplots}. 

\subsection{Qualitative validation of motif-based morphology}
\begin{figure*}
    \centering
    \begin{subfigure}[t]{0.83\textwidth}
        \centering
        \includegraphics[width=0.83\linewidth]{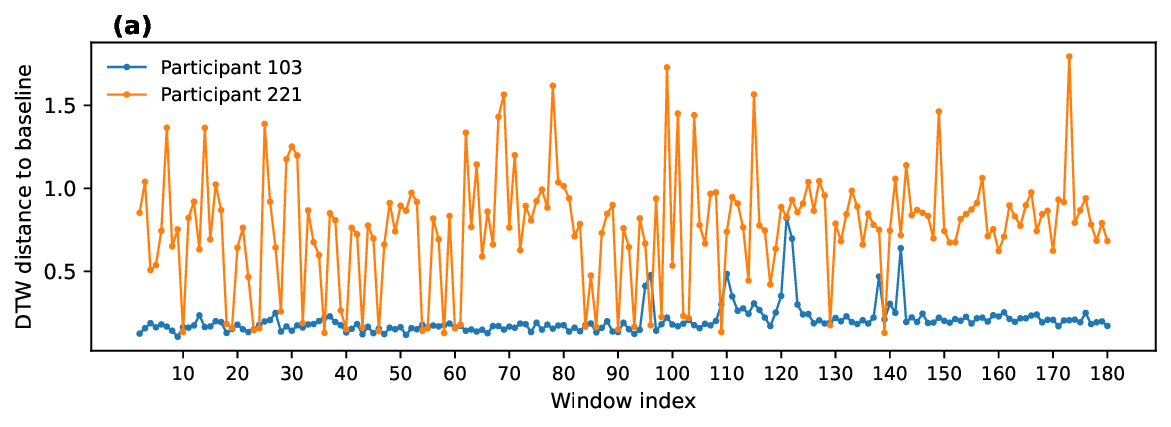}
        \label{fig:traj_baseline}
    \end{subfigure}
    \vspace{2pt}

      \begin{subfigure}[t]{0.83\textwidth}
        \centering
        \includegraphics[width=0.83\linewidth]{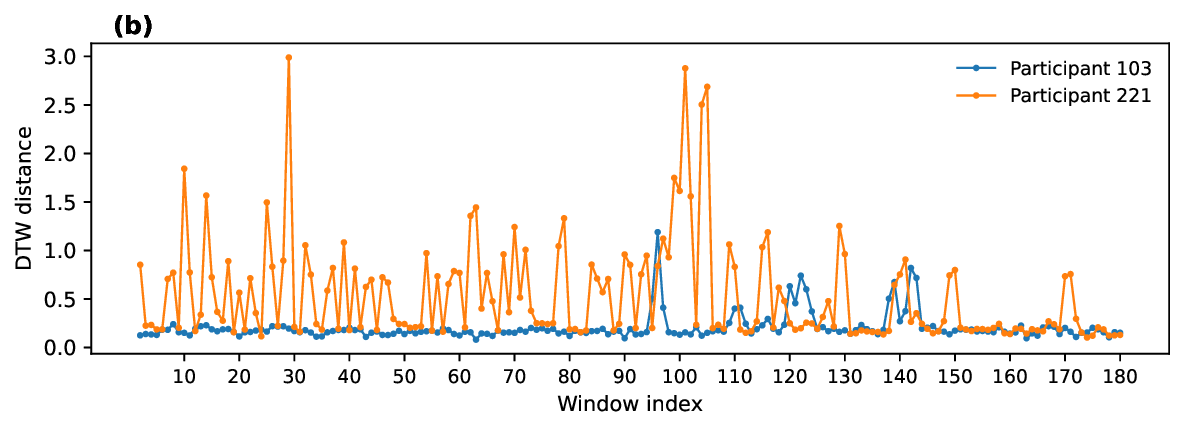}
        \label{fig:traj_prev}
    \end{subfigure}
    \caption{Motif distance trajectories over non-overlapping 10\,s windows in the MIT-BIH database for two participants. (a) Motif distances to personalised baseline motifs (PMDI). (b) Motif distances between consecutive windows (MII). Participant~103 (2/180 non-normal windows) exhibits stable, low-variance morphology. Participant~221 (163/180 non-normal windows) shows pronounced variability and frequent deviations, reflecting morphological instability due to frequent ventricular ectopy.}
    \label{fig:mitbih_trajectories}
\end{figure*}

Qualitative analysis illustrates how motif distance trajectories capture morphological variation in individual ECGs. Fig.~\ref{fig:mitbih_trajectories} shows trajectories for two contrasting participants: Participant~103 (male, 61~years), with a largely normal ECG (2/180 windows classified as non-normal), and Participant~221 (male, 83~years), with frequent ventricular ectopy, including PVCs with occasional couplets and short runs (163/180 windows classified as non-normal). Two trajectories are shown: motif deviations from each individual’s baseline motif (Fig.~\ref{fig:mitbih_trajectories}a), and motif deviations between consecutive windows (Fig.~\ref{fig:mitbih_trajectories}b). Normal recordings exhibit stable, low-variance trajectories, whereas abnormal ECGs display increased variability and pronounced deviations across windows.

\begin{figure*}
    \centering
    \begin{minipage}[t]{0.49\textwidth}
        \centering
        \vspace{0pt} 

        \begin{subfigure}[t]{\linewidth}
            \centering
            \includegraphics[width=\linewidth]{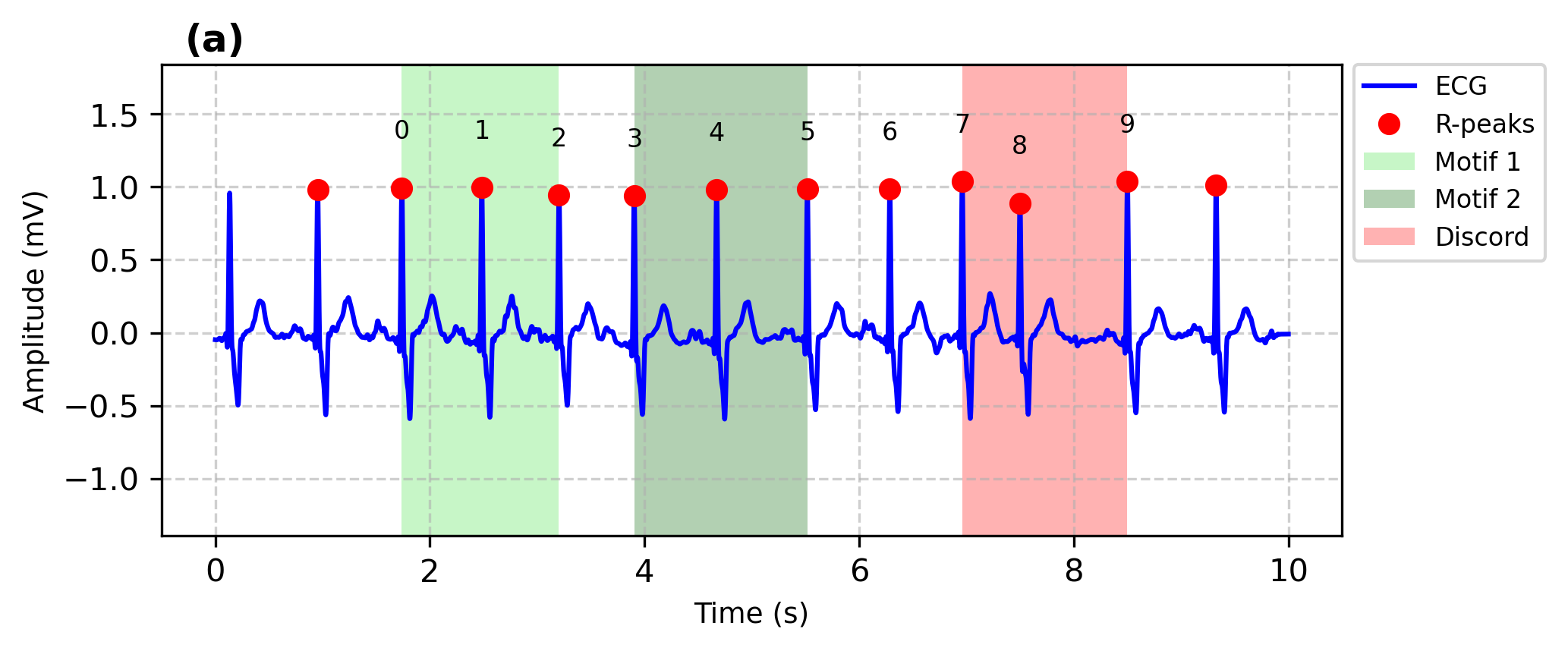}
            \label{fig:qual_ecg_window}
        \end{subfigure}

        \vspace{2mm}

        \begin{subfigure}[t]{\linewidth}
            \centering
            \includegraphics[width=\linewidth]{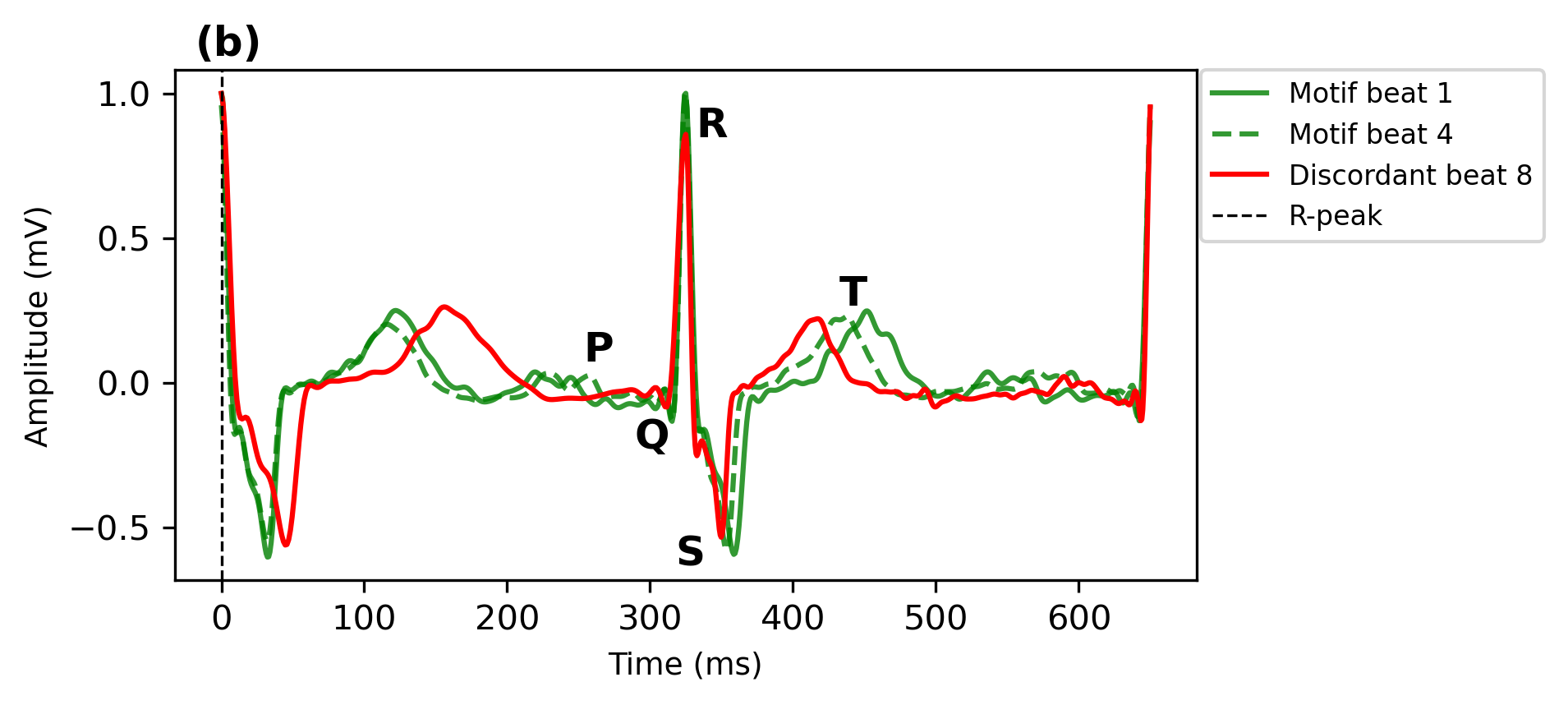}
            \label{fig:qual_waveforms}
        \end{subfigure}
    \end{minipage}
    \hfill
    \begin{minipage}[t]{0.49\textwidth}
        \centering
        \vspace{0pt} 

        \begin{subfigure}[t]{\linewidth}
            \centering
            \includegraphics[width=\linewidth]{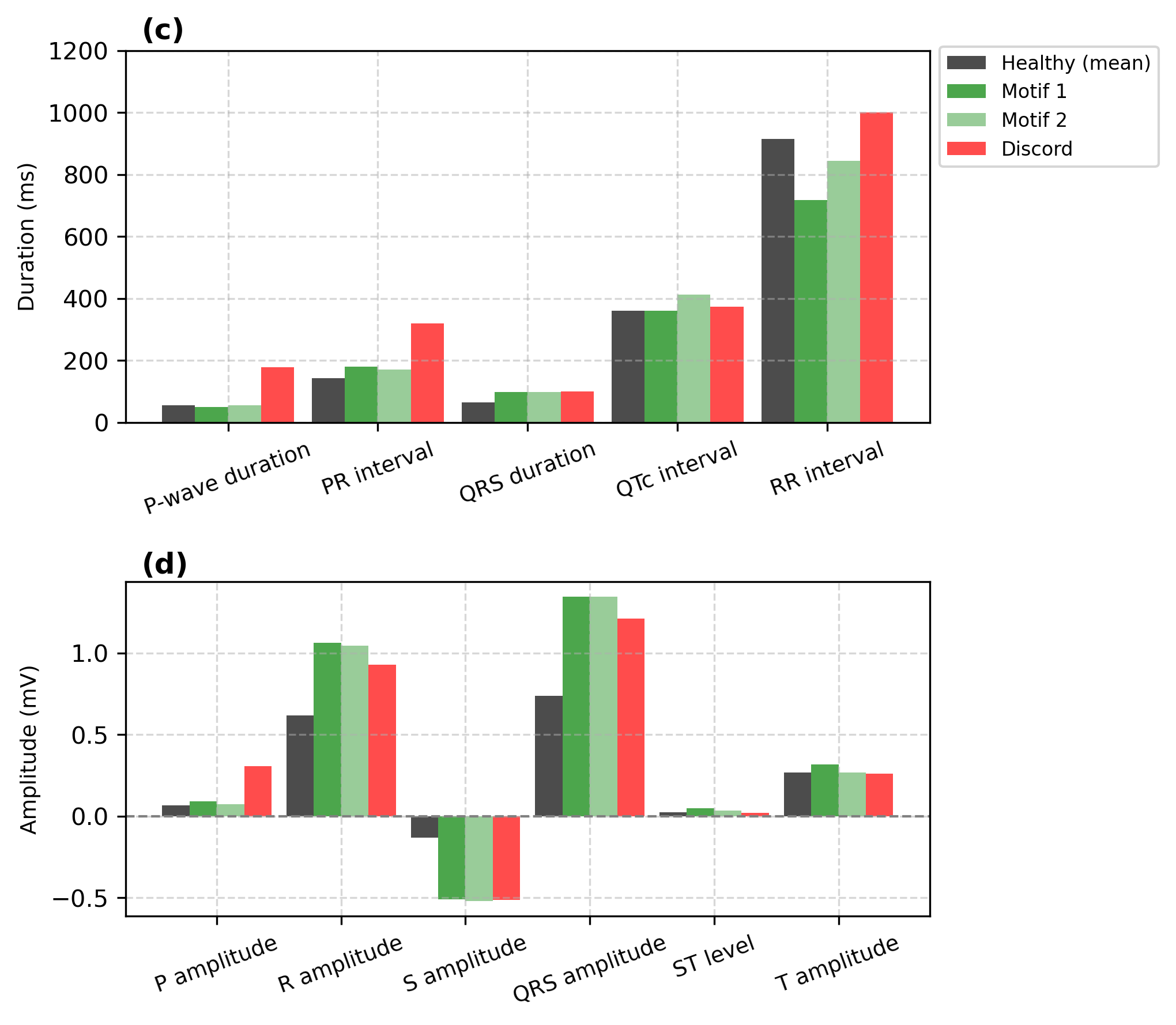}
            \label{fig:qual_morph_summary}
        \end{subfigure}
    \end{minipage}

    \vspace{1mm}
   \caption{Qualitative interpretation of motif- and discord-derived ECG morphology for PTB-XL recording 269 (female, 87 years) with sinus rhythm and conduction abnormalities (prolonged PR interval, left axis deviation, complete Right Bundle Branch Block) and Q waves consistent with prior inferior myocardial infarction. (a) Example 10\,s ECG window showing recurrent motifs (green) and discord (red). (b) Motif and discord waveforms, illustrating representative normal and atypical cardiac cycles. (c) Fiducial interval durations for motifs and discords relative to the NSR template, capturing temporal deviation from normal morphology. (d) Corresponding fiducial amplitudes relative to the NSR reference.}

    \label{fig:qualitative_panel}
\end{figure*}
\section{Discussion and Conclusion}
This work introduces a novel motif-based framework that defines beat-aligned ECG motifs as interpretable cardiac signatures and uses their temporal behaviour to quantify morphological drift and stability across short and long-duration ECG recordings. Unlike prior approaches based on generic subsequences, the proposed formulation is physiologically aligned to the cardiac cycle and enables interpretable tracking of morphology over time. In this study, motif-based deviation and instability metrics were evaluated as biomarkers of clinically meaningful departure from normal cardiac morphology using a single-lead ECG representation.

In PTB-XL (Fig.~\ref{fig:PTB-XL results}), deviation from the NSR template motif consistently separated normal ECGs from multiple abnormal diagnostic subtypes. Conduction disturbance (CD) subtypes showed the largest shifts in motif distance, consistent with pronounced alterations in QRS morphology and conduction timing. Myocardial infarction (MI) subtypes also demonstrated pronounced deviations, reflecting persistent waveform distortions associated with this condition. In contrast, hypertrophy (HYP) and ST/T change (STTC) subtypes exhibited comparatively smaller, but still statistically significant, shifts aligning with their more heterogeneous and often subtler morphological manifestations. Effect sizes varied across subtypes within a diagnostic category. For example, LPFB (Left Posterior Fascicular Block) and LBBB (Left Bundle Branch Block) exhibited among the highest motif distances within the CD group, whereas atrioventricular block subtypes showed more moderate deviations. A similar gradient was observed within MI and STTC subtypes, suggesting that motif distance captures a spectrum of morphological severity rather than acting as a binary abnormality indicator. This graded behaviour supports the utility of motif-based metrics as continuous markers of morphological deviation. Within hypertrophy, LVH (Left Ventricular Hypertrophy), an important risk marker for cardiovascular events, demonstrated the most statistically significant and largest effect size relative to normal ECG morphology. 

In MIT-BIH (Fig.~\ref{fig:MIT BIH-motif deviation boxplots}), motif-based deviation and instability metrics significantly distinguished predominantly normal from abnormal recordings, demonstrating generalisation to long-duration ambulatory ECGs. Moderate effect sizes for personalised baseline deviation and short-term instability highlight the value of personalised motifs for monitoring in the presence of substantial inter-subject variability.

Qualitative visualisations demonstrated how the proposed metrics map to ECG morphological changes. Normal ECGs showed largely stable motif trajectories, whereas abnormal recordings exhibited increased variability, with deviations aligned to ectopic activity, conduction abnormalities, and waveform distortion (Fig.~\ref{fig:mitbih_trajectories}). Motif and discord comparisons revealed  changes in temporal and amplitude features consistent with known electrophysiological alterations (Fig.~\ref{fig:qualitative_panel}). Overall, motif-based representations offer a unified and clinically meaningful framework for analysing ECG morphology across datasets, temporal scales, and diagnostic categories.

This work provides evidence that motif-based representations offer compact and interpretable signatures of ECG morphology suitable for longitudinal analysis. By distilling recurring waveform structure into concise descriptors, the proposed framework enables efficient tracking of morphological change and identification of emerging deviation from typical cardiac patterns. Motif and discord analysis support transparent characterisation of dominant and extreme morphological behaviour within an ECG recording. Such representations have the potential to support scalable ECG pre-screening and monitoring workflows by prioritising recordings for expert review, particularly in settings where subtle change detection and efficient interpretation are critical, such as athlete screening.
\par\vspace{1mm}
\noindent\textbf{Limitations and future work.}
Motifs capture the dominant recurring pattern within a signal window, and may therefore be influenced by recurrent artifact in poor-quality recordings. While beat-level quality control reduces obvious noise, artifact-aware motif selection and window-level signal-quality assessment remain important areas for future work. With improved artifact handling, discord analysis can be further developed to characterise extreme morphological variation alongside dominant motif behaviour.

A further limitation is the use of a simple initial baseline derived from the first analysis window for continuous ECG recordings. More robust baseline estimation strategies, including longer stable segments and adaptive or self-learning approaches, remain to be explored.

In addition, ECG morphology varies across leads due to differences in electrode placement and viewing angle. While this study used a single-lead setting, the framework naturally extends to multi-lead ECGs by extracting and comparing motifs per lead or by constructing joint multi-lead representations.

Future work will address these limitations through artifact-aware motif selection, improved baseline estimation, and multi-lead representations. We will also validate motif-derived metrics in athlete and other at-risk populations and develop additional measures of morphological heterogeneity.
\balance
\bibliographystyle{IEEEtran}

\end{document}